\documentclass[prl,twocolumn,floatix,superscriptaddress,letterpaper]{revtex4}

\usepackage{amsfonts}
\usepackage{amsmath}
\usepackage{amssymb}
\usepackage{epsfig}
\usepackage{epstopdf}
\usepackage{graphicx}
\usepackage{color}
\usepackage{bbold}

\begin{document}
\title{Photonic implementation of Majorana-based Berry phases}

\author{Jin-Shi Xu}
\thanks{These authors contributed equally to this work.}
\affiliation{CAS Key Laboratory of Quantum Information, University of Science and Technology of China, Hefei 230026, People's Republic of China}
\affiliation{Synergetic Innovation Center of Quantum Information and Quantum Physics, University of Science and Technology of China, Hefei 230026, People's Republic of China}

\author{Kai Sun}
\thanks{These authors contributed equally to this work.}
\affiliation{CAS Key Laboratory of Quantum Information, University of Science and Technology of China, Hefei 230026, People's Republic of China}
\affiliation{Synergetic Innovation Center of Quantum Information and Quantum Physics, University of Science and Technology of China, Hefei 230026, People's Republic of China}

\author{Jiannis K. Pachos}
\affiliation{School of Physics and Astronomy, University of Leeds, Leeds LS2 9JT, United Kingdom}

\author{Yong-Jian Han}
\email{smhan@ustc.edu.cn}
\affiliation{CAS Key Laboratory of Quantum Information, University of Science and Technology of China, Hefei 230026, People's Republic of China}
\affiliation{Synergetic Innovation Center of Quantum Information and Quantum Physics, University of Science and Technology of China, Hefei 230026, People's Republic of China}

\author{Chuan-Feng Li}
\email{cfli@ustc.edu.cn}
\affiliation{CAS Key Laboratory of Quantum Information, University of Science and Technology of China, Hefei 230026, People's Republic of China}
\affiliation{Synergetic Innovation Center of Quantum Information and Quantum Physics, University of Science and Technology of China, Hefei 230026, People's Republic of China}

\author{Guang-Can Guo}
\affiliation{CAS Key Laboratory of Quantum Information, University of Science and Technology of China, Hefei 230026, People's Republic of China}
\affiliation{Synergetic Innovation Center of Quantum Information and Quantum Physics, University of Science and Technology of China, Hefei 230026, People's Republic of China}

\date{\today }% It is always \today, today,

\begin{abstract}

Geometric phases, generated by cyclic evolutions of quantum systems, offer an inspiring playground for advancing fundamental physics and technologies, alike. Intriguingly, the exotic statistics of anyons realised in physical systems can be interpreted as a topological version of geometric phases. However, non-Abelian statistics has not yet been demonstrated in the laboratory. Here we employ an all optical quantum system that simulates the statistical evolution of Majorana fermions. As a result we experimentally realise non-Abelian Berry phases with the topological characteristic that they are invariant under continues deformations of their control parameters. We implement a universal set of Majorana inspired gates by performing topological and non-topological evolutions and investigate their resilience against perturbative errors. Our photonic experiment, while it is not scalable, it suggests the intriguing possibility of experimentally simulating Majorana statistics with scalable technologies.

\end{abstract}

\maketitle

%\tableofcontents{}

%\noindent
{\bf Introduction}

The Berry phase is one of physics most intriguing concepts~\cite{Berry}. It inspired numerous investigations towards theoretical frontiers with its possible generalisations~\cite{Wilczek} as well as technological applications in quantum computation~\cite{PachosZanardi}. At the forefront of research in geometric evolutions is the controlled realisation of anyonic statistics in condensed matter systems~\cite{Wilczek1984, Nayak2008, Sarma2015}. This is manifested by the cyclic evolution of two anyonic quasiparticles braided around each other. The anyonic quasiparticles are deemed Abelian or non-Abelian depending on the possible geometric evolutions from the exchange being simple global phase factors or non-commuting unitaries, respectively. The statistical character of the exchange evolutions dictates that the resulting geometric phases are topologically robust. This robustness is a very desirable characteristic as it makes non-Abelian anyons a promising platform for fault-tolerant quantum computation~\cite{Nayak2008, Sarma2015, Brown, Pachos}. In the past decades, non-Abelian anyons have been extensively theorised in condensed matter systems~\cite{Sarma2005,Fu2008,Sau2010,Lutchyn2010}. The most promising direction for realising non-Abelian anyons is the investigation on Majorana zero modes (MZMs). There are already several positive signatures for the realisation of MZMs in the laboratory~\cite{Mourik2012, Deng2012, Rokhinson2012, Mebrahtu2013, Nadj-Perge2014, Lee2014, Xu2015, Sun2016,He2017}. Nevertheless, the experimental realisation of braiding operations is still a challenging open problem.

%\noindent
{\bf Results}

{\em Encoding of MZM geometric evolutions.}
Here, we report the experimental quantum simulation of four MZMs braiding evolutions encoded in an all-optical system~\cite{Alan2012}. The MZMs are supported at the endpoints of two Kitaev Chain Models (KCM) comprised of fermions. To perform the encoding we first transform the fermion system, with Hamiltonian $H_\text{KCM}$, to a spin-1/2 system with Hamiltonian $H_\text{spin}$, through a unitary Jordan-Wigner (JW) transformation, ${\cal U}_\text{JW}$~\cite{Jordan1928,Kitaev2008}. The spin system is then encoded in the spatial modes of single photons~\cite{Xu2016}.

Under the Jordan-Wigner transformation the local Hamiltonians are unitarily connected
\begin{equation}
H_\text{spin} = {\cal U}^{}_\text{JW} H_\text{KCM}{\cal U}^\dagger_\text{JW}.
\label{eqn:unita}
\end{equation}
As a result, the time evolutions of the KCM and the spin system are identical when written in their corresponding basis states. The geometric phases that correspond to the braiding of MZMs are particular cases of time evolutions that are cyclic and adiabatic. Hence, the photonic system can simulate the statistical evolution of four MZMs by simulating the corresponding spin system. The possibility to generate an equivalent quantum evolution is in complete alignment with the spirit of quantum simulation~\cite{Georgescu2014}. The unitary equivalence (\ref{eqn:unita}) between the KCM and the spin system guarantees that the Berry phase obtained by evolving $H_\text{spin}$ is non-Abelian and topological in nature. Our previous experiment simulated the exchange of two MZMs positioned at the endpoints of the same chain, thus realising a topological Abelian Berry phase~\cite{Xu2016}.

The topological character of the spin model results from the topological character of the KCM. In the latter model the topological invariance corresponds to invariance of the geometric evolution against perturbations which are local in position space. As the environment is assumed to act locally in space, the KCM is a promising candidate for performing fault-tolerant quantum computation. The unitary transformation ${\cal U}_\text{JW}$ inherits the spin model with topologically invariant geometric evolutions, but now with respect to perturbations that are local in the parametric space of the adiabatic evolution. As these perturbations are not necessarily local in the position space, they may not correspond to possible environmental errors in the spin system.
%Actually, only $\mathbb{Z}_2$ symmetric perturbations in the spin system correspond to real error in the fermion system due to its parity symmetry.
%, while spin errors like $\sigma_x$ do not correspond to any physical error in fermionic system.
{ 
In addition, in our photonic experiment the resulting non-Abelian geometric phases are insensitive of the exact timing of each controlled evolution when it is large enough. This is a highly desirable characteristic that facilitates the experimental realisation of the non-Abelian evolution with high fidelity.}

By experimentally simulating the braiding of different pairs of MZMs we can realise only Clifford gates~\cite{Bravyi2006}, such as the Hadamard gate, $\text{H}=\frac{1}{\sqrt{2}}\left(\begin{array}{cc}
1 & -1\\
1 & 1
\end{array}\right)$, the $(-{\pi\over 4})$-phase gate, $\text{R}=\left(\begin{array}{cc}
1 & 0\\
0 & -i
\end{array}\right)$,
which are not universal for quantum computation~\cite{Gottesman1998}.
 The inclusion of a non-Clifford gate, such as the ${\pi\over 8}$-phase gate, $\text{T}=\left(\begin{array}{cc}
1 & 0\\
0 & e^{i\pi/4}
\end{array}\right)$, can resolve this problem~\cite{Bonderson2010}. We experimentally simulate the ${\pi\over 8}$-phase gate by moving two MZMs at the same site and exposing them to a controlled local perturbation. We experimentally demonstrate that, unlike the H and the R topological gates, the ${\pi\over 8}$-phase gate is not immune to local perturbations in the Majorana system. Nevertheless, `magic state distillation'~\cite{Bravyi2005} can be used to produce error-corrected ${\pi\over 8}$-phase gates from noisy ones. When access to an arbitrary number of Kitaev chains is possible, two-qubit topological gates can be realised by employing the control procedures presented here.

\begin{figure}[t]
\begin{centering}
\includegraphics[width=1\columnwidth]{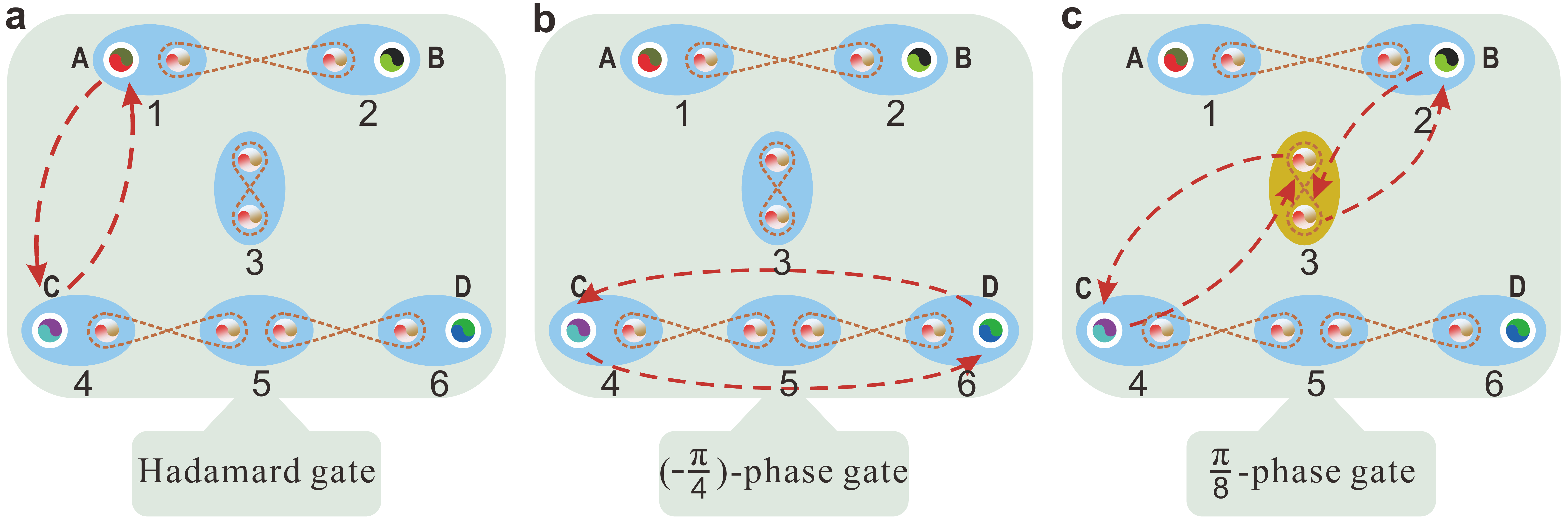}
\par\end{centering}
\protect\caption{The set of universal quantum gates. The Kitaev chains consist of six fermions (numbered from 1 to 6) with four endpoint Majorana zero modes A, B, C and D, which can be used to demonstrate the universal gates. Each two Majorana fermions in the blue ellipse form a conventional fermion. The dashed lines between different Majorana fermions represent the initial interactions between them. {\bf a.} The anticlockwise braiding of Majoranas A and C, implements a Hadamard gate, H, acting on the logical basis. {\bf b.} The anticlockwise braiding of Majoranas C and D, implements a $(-{\pi\over 4})$-phase gate, R, acting on the logical basis. {\bf c.} The real time population-dependent evolution on Majoranas B and C, which is realised by transporting the two MZMs to a single site (site 3 in our experiment) and applying a coupling between them, leads to a ${\pi\over 8}$-phase gate, T, acting on the logical basis.
}
\label{model}
\end{figure}

{\em Quantum gates based on Majorana braiding.} The smallest system of two connected Kitaev chains that remains fault-tolerant against local perturbations at all times during the braiding evolution comprises of six fermion sites~\cite{Kitaev2001}. Using six rather than five sites guarantees that no pairs of MZMs ever meet at the same site, which would render them unprotected to local perturbations. Here we describe these fermions through the canonical operators $c_j$ and $c_j^\dagger$, with positions $j=1,...,6$, where $j=1, 2$ constitutes the first chain, $j=4, 5, 6$ constitutes the second and $j=3$ corresponds to the link between them, as shown in Fig.~\ref{model}.
%%%%%%%%FIGURE 1%%%%%%%%%%%%%%
%%%%% FIGURE %%%%%
The Kitaev model for the two chains is given in terms of Majorana operators, $\gamma_{ja}=c_{j}+c^{\dag}_{j}$ and $\gamma_{jb}=i(c^{\dag}_{j}-c_{j})$, by the Hamiltonian
\begin{equation}
 H_{M_0}=i(\gamma_{1b} \gamma_{2a}+\gamma_{4b} \gamma_{5a}+\gamma_{5b} \gamma_{6a})+i\gamma_{3a} \gamma_{3b}.
 \label{eqn:Ham0}
\end{equation}
The Majorana operators $\gamma_m$ satisfy the relations $\gamma^{\dag}_m=\gamma_m$ and $\gamma_l\gamma_m+\gamma_m\gamma_l=2\delta_{lm}$ for $l,m=1a,1b,...,6a,6b$. Note that the particular operators $\gamma_{1a}$, $\gamma_{2b}$, $\gamma_{4a}$ and $\gamma_{6b}$ are not present in Hamiltonian (\ref{eqn:Ham0}), so $[H_{M_0}, \gamma_j]=0$ for $j=1a,2b,4a,6b$. As a result, these Majorana modes have zero energy, giving rise to four endpoint MZMs, which we denote as A, B, C and D in Fig.~\ref{model}. The logical qubit states are taken to be $|0_L\rangle = |00_g\rangle$ and $|1_L\rangle = |11_g\rangle$ corresponding to the degenerate ground-states of $H_{M_0}$ with even fermion parity, given by $|00_g\rangle=Nf_1d_1f_2d_2d_3|\text{vac}\rangle$ and $|11_g\rangle=f^{\dag}_1f^{\dag}_2|00_g\rangle$, where $f_1=(\gamma_{1a}+i\gamma_{2b})/2$, $f_2=(\gamma_{6b}+i\gamma_{4a})/2$, $d_1=(\gamma_{1b}+i\gamma_{2a})/2$, $d_2=(\gamma_{4b}+i\gamma_{5a})/2$ and $d_3=(\gamma_{5b}+i\gamma_{6a})/2$. For convenience we denote the appropriate normalisation constant by $N$.

The Hadamard gate $\text{H}$ on the logical qubit can be realised by anticlockwise braiding the MZMs A and C positioned at sites 1 and 4, respectively, as shown in Fig.~\ref{model}a.
The transport of the MZMs around the chain network is performed by adiabatically evolving the system through the following sequence of Hamiltonians, $H_{M_0}$, $H_{h_1}$, $H_{h_2}$, $H_{h_3}$ and $H_{M_0}$, where
\begin{equation}
\begin{split}
  &H_{h_1}=i(\gamma_{1b} \gamma_{2a}+\gamma_{1a} \gamma_{3a}+\gamma_{5b} \gamma_{6a})+i\gamma_{4a} \gamma_{4b}, \\
  &H_{h_2}=i(\gamma_{1b} \gamma_{2a}+\gamma_{1a} \gamma_{3a}+\gamma_{3b} \gamma_{4b}+\gamma_{5b} \gamma_{6a}), \\
  &H_{h_3}=i(\gamma_{1b} \gamma_{2a}+\gamma_{1a} \gamma_{3a}+\gamma_{4b} \gamma_{5a}+\gamma_{5b} \gamma_{6a}).
\end{split}
\label{eqn:fermham}
\end{equation}
A depiction of the resulting MZMs transportation is shown in the Section I.B in Supplementary Material (SM). The ground states of these Hamiltonians have the MZMs located at the desired sites. Hence, braiding can be implemented by a set of consecutive imaginary-time evolution (ITE) operators, $e^{-H_{M_0}t}$, $e^{-H_{h_1}t}$, $e^{-H_{h_2}t}$, $e^{-H_{h_3}t}$ and $e^{-H_{M_0}t}$, where $t$ is taken to be large enough for these operators to faithfully represent projectors onto the corresponding ground states up to overall normalisation~\cite{Xu2016}. Due to the topological nature of the produced evolutions, the exact value of $t$ does not matter as long as it is long enough to suppress the contribution from the excited states (see Materials and Methods and Section IA in SM). The theoretically expected non-Abelian Berry phase resulting from this set of evolutions is given by
$\text{H}={1 \over \sqrt{2}}\left(
   \begin{array}{cc}
     1 & -1 \\
     1 & 1 \\
   \end{array}
 \right)$, when written in the logical basis $\{|00_{g}\rangle, |11_{g}\rangle\}$ (see Materials and Methods).

%%%%% FIGURE 2 %%%%%
\begin{figure*}[tbph]
\includegraphics[width=1.8\columnwidth]{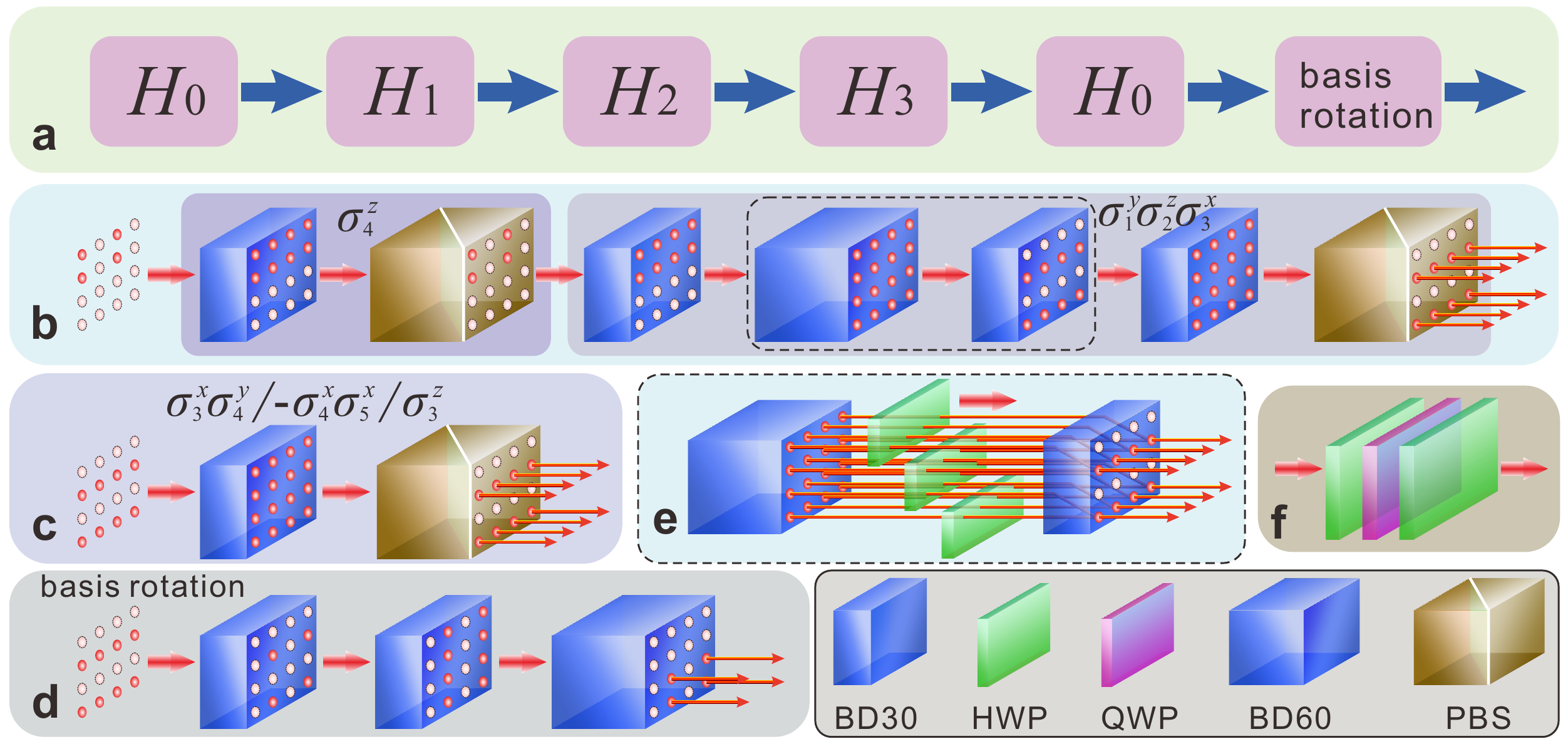}
\caption{Experimental setup. {\bf a.} The imaginary-time evolutions (ITEs) between Hamiltonians that exchange Majorana zero modes A and C. {\bf b.} The setup to realise the ITE of $H_{1}$ (needs only $\sigma_{4}^{z}$ and $\sigma_{1}^{y}\sigma_{2}^{z}\sigma_{3}^{x}$). The state is initially prepared to be the ground state of $H_{0}$ involving four spatial modes, represented by solid circles. After rotated by beam displacers (BD30 with beams separated by 3.0 mm and BD60 with beams separated by 6.0 mm), half-wave plates (HWPs) and quarter-wave plates (QWPs), and subsequently dissipated by polarisation beam splitters (PBSs), there are eight output spatial modes. One of the operation processes represented by the arrows is shown in {\bf e} with HWPs set at different angles operating on different spatial modes. The combination of HWPs and a QWP in {\bf f} is used to exchange basis between the Pauli operators $\sigma_{y}$ and $\sigma_{x}$ ($\sigma_{z}$). {\bf c.} The setup for the ITE of $H_{2}$ (needs only $\sigma_{3}^{x}\sigma_{4}^{y}$). The subsequent ITEs of $H_{3}$ (needs only $-\sigma_{4}^{x}\sigma_{5}^{x}$) and $H_{0}$ (needs only $\sigma_{3}^{z}$) are the same as $H_{2}$. {\bf d.} The setup for basis rotation is used to rotate the output state along the same basis as the input state.}
\label{setup}
\end{figure*}
%%%%% FIGURE 2 %%%%%

To realise the $\text{R}$ gate on the logical qubit we need to anticlockwise braid the MZMs C and D, as shown in Fig.~\ref{model}b. The experimental simulation of the braiding evolution is performed by switching between the corresponding Majorana Hamiltonians $H_{r_1},H_{r_2}$ and $H_{r_3}$. This time evolution can be implemented by a set of consecutive ITE operators, $e^{-H_{M_0}t}$, $e^{-H_{r_1}t}$, $e^{-H_{r_2}t}$, $e^{-H_{r_3}t}$ and $e^{-H_{M_0}t}$. The detailed process is given in Section I.C of SM. The resulting non-Abelian Berry phase is given by
$\text{R}=\left(
   \begin{array}{cc}
     1 & 0 \\
     0 & -i \\
   \end{array}
 \right)$ in the logical basis. The corresponding braiding with a single chain was realised in~\cite{Xu2016}. The  Hermitian conjugate gates, H$^\dagger$ and R$^\dagger$, are produced by reversing the orientation of the exchanging paths.
Realising the Hadamard gate, $\text{H}$, and the $(-{\pi\over 4})$-phase gate, $\text{R}$, by braiding MZMs demonstrates the non-Abelian character of the generated Berry phases. When these two operations are performed in reverse order, they give a different composite geometric evolution, since $\text{HR}\neq \text{RH}$.

To realise the ${\pi\over 8}$-phase gate we place two MZMs at the same site and apply a local field. This causes the splitting of the ground state degeneracy for a certain time, during which the appropriate dynamical phase factor is accumulated~\cite{Bonderson2010}. In particular, we transport the B and C MZMs to site 3 by a set of ITE operations. Then the population dependent Hamiltonian $H_e=-i\gamma_{3a}\gamma_{3b}$ is operated for a certain time $\tau$, as shown in Fig.~\ref{model}c. Finally, the MZMs are transferred back to their initial position. The details of this process can be found in Section I.D of SM. During this evolution, the qubit states are transformed by
$\text{M}=\left(
   \begin{array}{cc}
     \cos\tau & -i\sin\tau \\
     -i\sin\tau & \cos\tau \\
   \end{array}
 \right) = e^{-i\sigma_x\tau}.
$
With the help of the Hadamard gate we can obtain the ${\pi\over 8}$-phase gate as $\text{H}^\dagger \text{M}\text{H}=e^{-i\tau}
\left(
   \begin{array}{cc}
     1 & 0 \\
     0 & e^{2i\tau } \\
   \end{array}
 \right)$, by choosing the time to be $\tau={\pi\over 8}$. This gate is not protected against noise perturbations acting on site 3 when both MZMs are positioned there. Moreover, unlike the braiding gates, the dynamical gate is sensitive to timing errors.

{\em {Spin encoding of two-chain system.}} To experimentally simulate the braiding evolutions of MZMs A and C, we transform the fermionic Hamiltonians $H_{M_0}$, $H_{h_1}$, $H_{h_2}$ and $H_{h_3}$ of (\ref{eqn:Ham0}) and  (\ref{eqn:fermham}), via a JW transformation, into the equivalent spin Hamiltonians, $H_{0}$, $H_{1}$, $H_{2}$ and $H_{3}$, respectively, where
\begin{equation}
\begin{split}
  &H_{0} =-\sigma_{1}^{x}\sigma_{2}^{x}+\sigma_{3}^{z}-\sigma_{4}^{x}\sigma_{5}^{x} -\sigma_{5}^{x}\sigma_{6}^{x}, \\
  &H_{1}=-\sigma_{1}^{x}\sigma_{2}^{x}+\sigma_{1}^{y}\sigma_{2}^{z}\sigma_{3}^{x}+\sigma_{4}^{z}-\sigma_{5}^{x}\sigma_{6}^{x}, \\
  &H_{2}=-\sigma_{1}^{x}\sigma_{2}^{x}+\sigma_{1}^{y}\sigma_{2}^{z}\sigma_{3}^{x}+\sigma_{3}^{x}\sigma_{4}^{y}-\sigma_{5}^{x}\sigma_{6}^{x}, \\
  &H_{3}=-\sigma_{1}^{x}\sigma_{2}^{x}+\sigma_{1}^{y}\sigma_{2}^{z}\sigma_{3}^{x}-\sigma_{4}^{x}\sigma_{5}^{x}-\sigma_{5}^{x}\sigma_{6}^{x}.
\end{split}
\label{eqn:spinHam}
\end{equation}
During the adiabatic process the spin system has the same spectrum as the fermion system at all times. Hence, both systems share the same time evolution operators when written in their corresponding basis~\cite{Xu2016}. In particular, the non-Abelian geometric phase obtained during the transport of MZMs can be faithfully studied in the equivalent spin system. Due to the commutation relations between the terms of $H_{0}$, $H_{1}$, $H_{2}$ and $H_{3}$, the total process of ITE can be further simplified to be $e^{-H_{0}t}e^{-H_{3}t}e^{-H_{2}t}e^{-H_{1}t}|\phi_{0}\rangle=e^{-\sigma_{3}^{z}t}e^{\sigma_{4}^{x}\sigma_{5}^{x}t}e^{-\sigma_{3}^{x}\sigma_{4}^{y}t} e^{-\sigma_{1}^{y}\sigma_{2}^{z}\sigma_{3}^{x}t}e^{-\sigma_{4}^{z}t}|\phi_{0}\rangle$, where $|\phi_{0}\rangle$ is the ground state of $H_0$. To experimentally simulate the above dynamics, we need, in principle, a $2^7$-dimensional Hilbert space, that corresponds to six spins for the chain network and an extra spin for implementing dissipation. However, due to the character of the ITE, we need to focus only on manipulations that act on the low-energy subspace, which is $2^5$-dimensional (see Materials and Methods). While our photonic simulator has limited scalability as it does not possess a tensor product structure, we successfully managed to encode the full low-energy Hilbert space.

The experimental setup that realises the adiabatic evolutions between the spin Hamiltonians (\ref{eqn:spinHam}) and, as a consequence, the evolutions that correspond to braiding MZMs A and C is shown in Fig.~\ref{setup}. We encode the quantum states in the optical spatial modes of photons and manipulate them by beam displacers (BDs). A beam displacer is a birefringent crystal, which separates light beams with horizontal and vertical polarisations by a certain displacement that depends on the length of the crystal~\cite{Kitagawa2012}. In our experiment, the polarisation of the photons is used as the environmental degree of freedom for the realisation of the ITE operations. The coupling between the spatial modes and the photon polarisation is achieved using half-wave plates (HWPs) and quarter-wave plates (QWPs), which rotate the polarisation of the corresponding modes. A dissipative evolution is accomplished in two steps. Initially, photons are passed through a polarising beam splitter (PBS), which transmits the horizontal component and reflects the vertical one. Subsequently, photons with vertical polarisation are completely dissipated and only the ones with horizontal polarisation are preserved. The resulting states correspond to the ground state of the spin chain system. In this way, the state $|\phi_{0}\rangle$ is initially prepared and is then sent to the ITE operation of $H_{1}$, $H_{2}$, $H_{3}$ and $H_{0}$ for the braiding of A and C with the dynamical map shown in Fig.~\ref{setup}a. The ITE operations are realised in Figs.~\ref{setup}b (see Materials and Methods) and c with one of the detailed processes shown in Fig.~\ref{setup}e. The combination of HWPs and a QWP in Fig.~\ref{setup}f is used to exchange basis between Pauli operators $\sigma_{y}$ and $\sigma_{x}$ ($\sigma_{z}$). The setup of basis rotation shown in Fig.~\ref{setup}d is used to rotate the output state onto the same basis of the input state. During the experiment, we need to construct a stable interferometer with sixteen spatial modes. The relative phases in the interferometer are compensated by inserting thin glasses in the corresponding paths (not shown in Fig. 2). The effective operator of our setup (with four input modes and four output modes) is reconstructed by quantum process tomography with 256 measurements~\cite{OBrien2004}. The experimental configurations that demonstrate the braiding of C and D and the ${\pi\over 8}$-phase gate are similar to the one shown in Fig.~\ref{setup}a and are given in Sections II.A and II.B of SM. The cross section images for the state evolution during the ITE operation are shown in Section I.I of SM.

\begin{figure}[t]
\begin{centering}
\includegraphics[width=1\columnwidth]{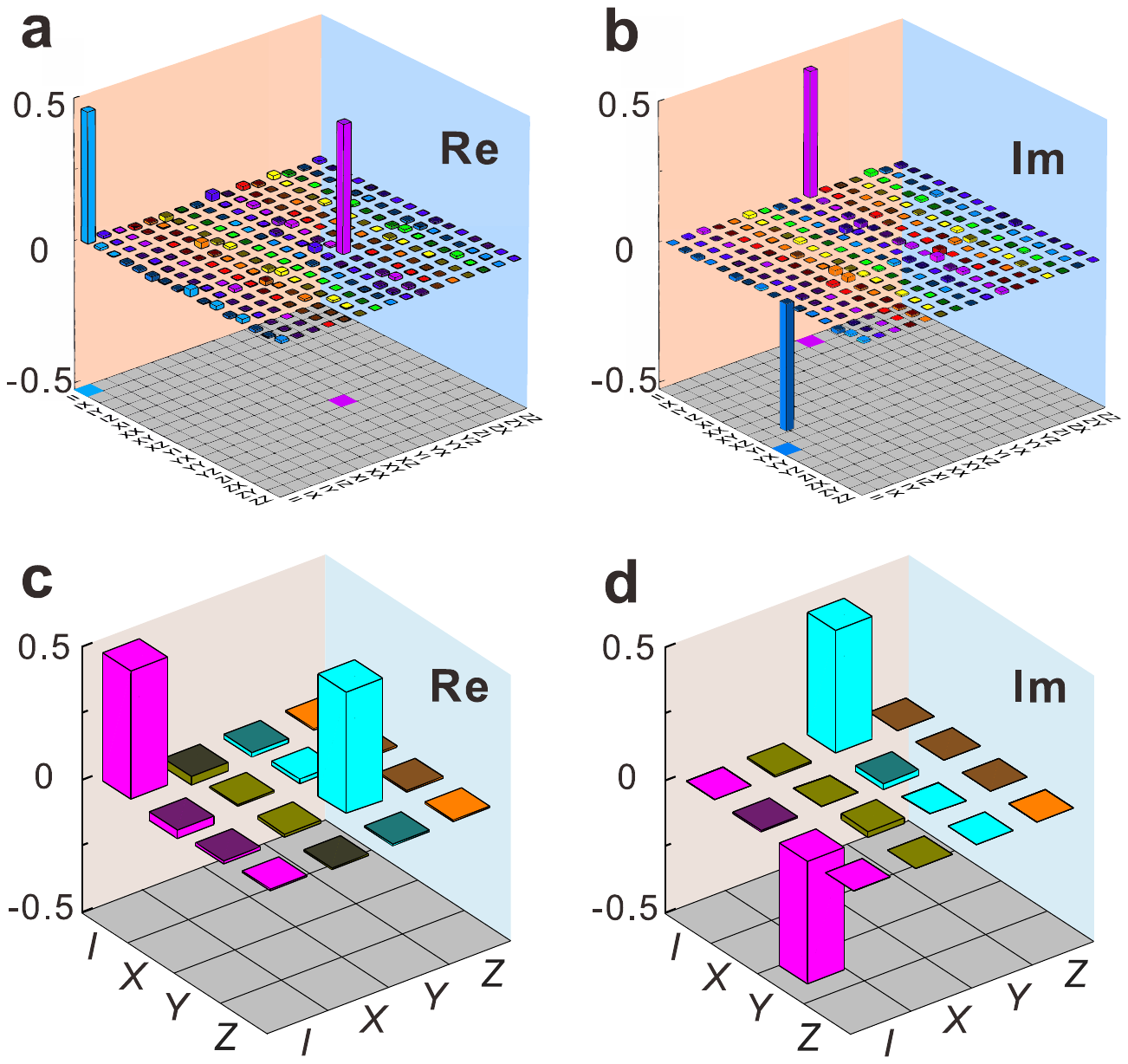}
\par\end{centering}
\protect\caption{ Experimentally obtained density matrices for the Hadamard gate operation. {\bf a.} Real (Re) and {\bf b.} Imaginary (Im) parts of the Hadamard gate operator in the basis $\{|00_{g}\rangle, |01_{g}\rangle, |10_{g}\rangle, |11_{g}\rangle\}$. {\bf c.} Real (Re) and {\bf d.} Imaginary (Im) parts of the Hadamard gate operator in the computational basis $\{|00_{g}\rangle, |11_{g}\rangle\}$. The measurement operators I, X, Y and Z represent the identity, $\sigma^{x}$, $\sigma^{y}$ and $\sigma^{z}$, respectively.}
\label{density}
\end{figure}

\begin{figure}[t]
\begin{centering}
\includegraphics[width=1\columnwidth]{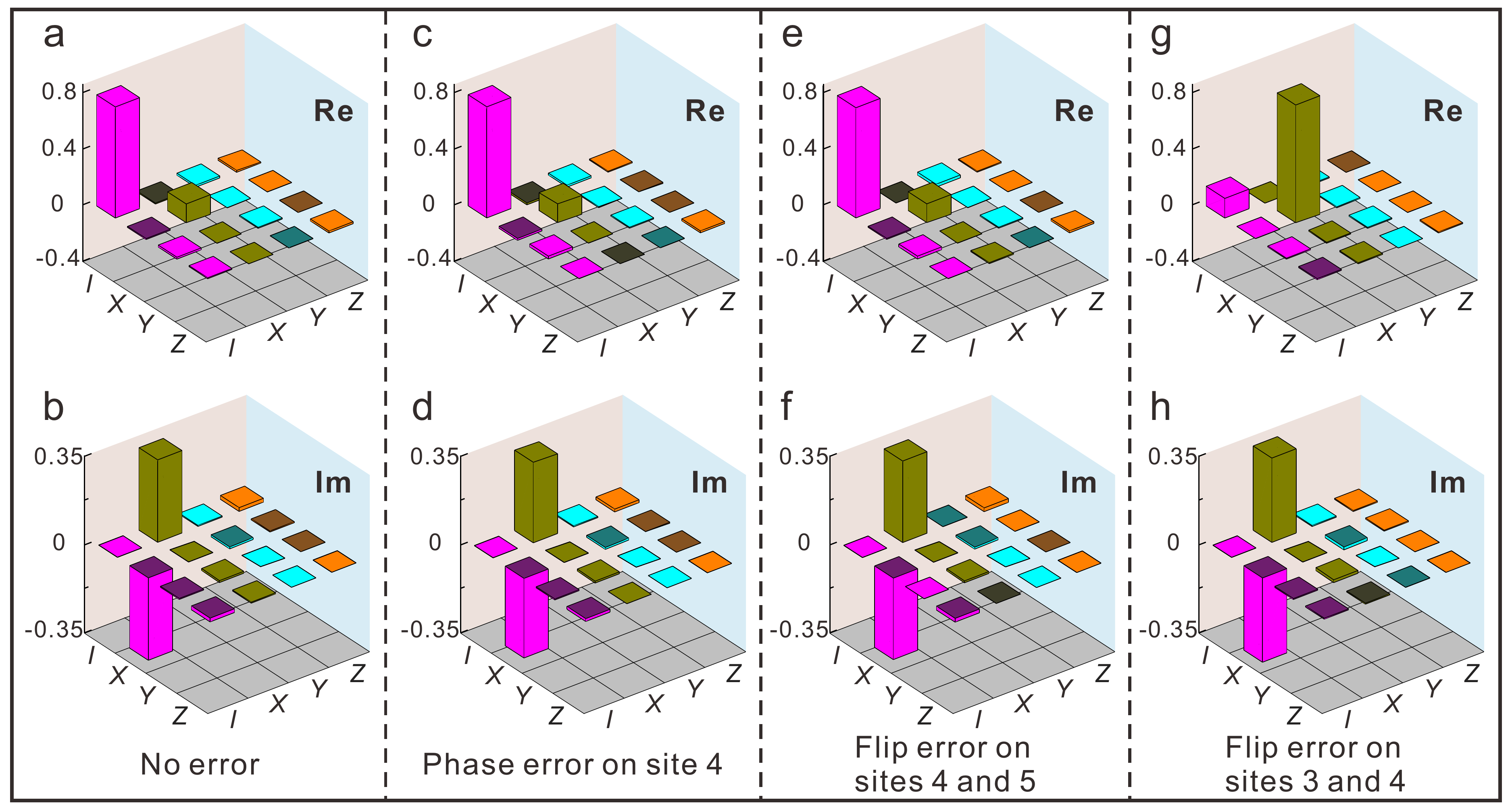}
\par\end{centering}
\protect\caption{Experimental results of the effect phase and flip errors have during the ${\pi\over 8}$-phase gate on the logical qubit encoded in the basis $|00_{g}\rangle$ and $|11_{g}\rangle$. {\bf a.} Real (Re) and {\bf b.} Imaginary (Im) parts of the density matrix without errors. {\bf c.} Real (Re) and {\bf d.} Imaginary (Im) parts of the density matrix with phase error on site 4. {\bf e.} Real (Re) and {\bf f.} Imaginary (Im) parts of the density matrix with flip error on sites 4 and 5. {\bf g.} Real (Re) and {\bf h.} Imaginary (Im) parts of the density matrix with flip error on sites 3 and 4.}
\label{noises}
\end{figure}

{\em Realisation of quantum gates.} In order to characterise the quantum gates resulting from the braiding of MZMs, we experimentally reconstruct the whole density matrix in the basis of  $\{|00_{g}\rangle, |01_{g}\rangle, |10_{g}\rangle, |11_{g}\rangle\}$ (see Materials and Methods). The operators can be described in the 16-dimensional basis spanned by $\Sigma^\alpha\otimes \Sigma^\beta$ with $\Sigma^{\alpha(\beta)}$ being I, X, Y, Z for $\alpha (\beta)=0,1,2,3$, respectively, corresponding to the identity matrix and the three Pauli operators.
%$\{{\rm II}, {\rm IX}, {\rm IY}, {\rm IZ}, {\rm XI}, {\rm XX}, {\rm XY}, {\rm XZ}, {\rm YI}, {\rm YX}, {\rm YY}, {\rm YZ}, {\rm ZI}, {\rm ZX}, {\rm ZY}, {\rm ZZ}\}$, where I represents the identity and where X, Y and Z represent the three Pauli operators $\sigma^{x}$, $\sigma^{y}$ and $\sigma^{z}$, respectively.
The experimental result is shown in Figs.~\ref{density}a (real part) and b (imaginary part). The evolution corresponds to a Hadamard gate acting on the Majorana-based encoding qubit. To clearly show this, we express the data in the logical basis $\{|00_{g}\rangle, |11_{g}\rangle\}$. The result for the corresponding implementation of ${\rm H}=(\text{I}-i\text{Y})/\sqrt{2}$ is shown in Figs.~\ref{density}c (real part) and d (imaginary part). The overall fidelity of the Hadamard operator is $93.47\pm0.02\%$.
For the ${\pi\over 8}$-phase gate,  ${\rm T}=\cos{\pi\over 8}\text{I}-i\sin{\pi\over 8}\text{X}$, the experimental fidelity is $92.57\pm0.01\%$. The real and imaginary parts of the density matrix are shown in Fig.~\ref{noises}a and b. The $(-{\pi \over 4})$-phase gate, $\text{R} = \cos {\pi\over 4}\text{I} + i\sin{\pi\over 4}\text{Z}$, is further demonstrated with a fidelity of $93.44\pm0.01\%$. All the density matrices corresponding to these operations are given in Section II.D in SM. The uncertainty in the fidelities is deduced from the Poissonian photon counting noise~\cite{Xu2016}.

{  In our photonic experimental system, the main naturally occurring errors include the imperfect interference, the rotation errors of the wave plates and the photon statistics fluctuation from the source. These errors are well under control and they lead to the reduction of the fidelity ($\sim$93\%). On the other hand, these fidelities are by large independent on the exact value of the imaginary time evolution parameter $t$ as long as it is large enough to suppress contributions from excited states. Indeed, our numerical simulations show that the resulting evolutions stay unaffected even if we increase $t$ by a factor of 2. In our experiment, the timing $t$ depends on the ratio between the reflected and transmitted parts of the vertical polarisation after the PBS in our experiment, which can be higher than 500:1.
}

%%%%% FIGURE 3 %%%%%
%%%%% FIGURE %%%%%

{\em Fault-tolerance.} During the realisation of the H and R topological gates the braided MZMs are never positioned at the same site. So these gates are immune to arbitrary single-site perturbative errors in the Majorana system~\cite{Xu2016}. The ${\pi\over 8}$-phase gate is not expected to be resilient against perturbations that act on site 3, where the two MZMs are brought together. Such perturbations can lift the degeneracy of the logical basis states thereby causing dephasing of the encoded quantum information.

In our experiment, the ITE operators not only drive the evolutions that result to quantum gates, they also induce the effective interaction to supply the protection of the system. To experimentally probe this behaviour, we add phase errors in the MZM system, realised by the spin operation $(1+\sigma^{z})/2$, acting on various sites during the control operations on the spin chains that give the ${\pi\over 8}$-phase gate. The experimental setup is given in Section II.B of SM. The effective one-qubit gates in our scheme act on the space spanned by $\{|00_{g}\rangle,|11_{g}\rangle\}$. Fig.~\ref{noises} shows the final experimental density matrices with errors on different sites.  For comparison, Figs.~\ref{noises}a and b show the real and imaginary parts of the density matrix after the implementation of the ${\pi\over 8}$-phase gate without adding any errors at all. When local phase errors happen on site 4 during the gate manipulations, only one MZM is disturbed at a time and the operation remains unaffected. This resilience of the encoded information is clearly shown in Figs.~\ref{noises}c and d. On the other hand, when the phase error is implemented on site 3, both MZMs are simultaneously disturbed. Hence, the final state is corrupted as the evolution is not topologically protected. A detailed analysis is given in Sections I.E and I.F of SM.

%%%%% FIGURE 4 %%%%%
%%%%% FIGURE %%%%%

Besides phase errors, we also consider flip errors. In the fermionic system a flip error happens when a fermion erroneously tunnels between neighbouring sites of the wire~\cite{Kitaev2001}. This evolution can be exponentially suppressed by increasing the potential barrier between the two sites. In the spin system, flip errors are realised by $(\sigma^{y}\sigma^{y}+\sigma^{x}\sigma^{x})/2$. These errors degrade the encoded information, only if the MZMs are positioned on the same or on neighbouring sites to where the flip error acts. To demonstrate this, we implement a flip error between sites 4 and 5 when the MZMs are both on site 3. In this case the operation remains unchanged, as shown in Figs.~\ref{noises}e and f. However, if the flip error acts on sites 3 and 4 while both MZMs are positioned on site 3, then the operation is corrupted, as shown in Figs.~\ref{noises}g and h. The theoretical analysis can be found in Sections I.G and I.H of SM. {  Apart from the phase and flip errors that have their origin in the MZM system, the geometric phases are actually protected against all noise with $\mathbb{Z}_2$ symmetry of the spin system.}

%\noindent
{\bf Discussion}

In summary, we have experimentally demonstrated that it is in principle possible to implement non-Abelian Berry phases that simulate fault-tolerant quantum computation with MZMs. Our experiment is based on the dissipation method for the implementation of Berry phases introduced in our previous work~\cite{Xu2016}. There we experimentally simulate the evolutions of a single Kitaev chain corresponding to the exchange of its two endpoint MZMs and demonstrate the topological invariance of the resulting Abelian Berry phases. Here we experimentally simulate the evolutions of two chains with four endpoint MZMs. This setup allows us to generate with high accuracy Berry phases that are both non-Abelian and topological in nature, mirroring the braiding statistics of MZMs. With these evolutions in hand, we can implement several quantum algorithms topologically, such as the Deutsch-Jozsa algorithm~\cite{Deutsch1992}. A detail protocol is given in Section I.I of SM. While our work simulates the {\em evolution operator} of Majorana braiding, the physical system we use is not the same, but unitarily equivalent to that of Majorana fermions. In our simulation the obtained geometric phases are invariant under continuous variations of the control parameters, as is the case with the Majorana braiding. This invariance is of importance to quantum computation applications, as it provides stability against control errors of the experimental parameters.

When more than two Kitaev chains can be encoded, topological quantum computation with MZMs can be simulated by employing exactly the same control procedures demonstrated here, applied to arbitrary pairs of chains.
Due to the specific nature of our optical experiment we are able to perform control operations with very high fidelity, but the scalability of our system is limited. Scalable MZM quantum computation can be experimentally simulated by translating our photonic simulator implementation to scalable systems, such as ion traps~\cite{Barreiro2011}, ultracold atoms~\cite{Daley2014} and superconducting circuits~\cite{You2015} technologies, where the ITE dissipation methods have already been established.

%\noindent
{\bf Materials and Methods}

{\em Performing imaginary-time evolution.}
Any pure state $|\phi\rangle$ can be expressed in a complete set of eigenstates $|e_{k}\rangle$ of a certain Hamiltonian $H$ as $|\phi\rangle=\sum_{k}q_{k}|e_{k}\rangle$, where $q_{k}$'s represent the corresponding complex amplitudes. The imaginary-time evolution (ITE) operator associated to $H$ is given by $\exp(-Ht)\sum_{k}q_{k}|e_{k}\rangle=\sum_{k}q_{k}\exp(-E_{k} t)|e_{k}\rangle$, where $E_{k}$ is the eigenvalue corresponding to~$|e_{k}\rangle$. After the ITE, the amplitude $q_{k}$ is changed to $q_{k}\exp(-E_{k} t)$. The decay of the amplitude dependents exponentially on the energy: the higher the energy, the faster the decay of the amplitude. Therefore, for sufficiently large $t$ only the ground state of $H$ (with lowest energy) survives with high fidelity.

The implementation of the ITE operations, employed to perform the braiding, can be simplified as the terms of the corresponding Hamiltonians commute with each other. For example, $e^{-H_{0}t}$ can be decomposed into $e^{\sigma_{5}^{x}\sigma_{6}^{x}t}e^{-\sigma_{3}^{z}t}e^{\sigma_{4}^{x}\sigma_{5}^{x}t}e^{\sigma_{1}^{x}\sigma_{2}^{x}t}$. The ITE operator of each term can be directly implemented by local unitary operations and dissipation. To perform the dissipation in a controlled way, an environmental degree of freedom is introduced, which is appropriately coupled to the system. The total state of the system and its environment can be written as $|\phi_{t}\rangle=(|\phi_{g}\rangle|0_{e}\rangle+|\phi_{g}^{\perp}\rangle|1_{e}\rangle)/\sqrt{2}$, where $|\phi_{g}^{\perp}\rangle$ denotes the states that are orthogonal to the ground state $|\phi_{g}\rangle$ of the system. The environmental state $|1_{e}\rangle$ is dissipated during the evolution, and only $|0_{e}\rangle$ is preserved. Therefore, the ground state of the corresponding Hamiltonian is obtained.

{\em Experimental procedure for implementing the ITE of $H_{1}$.} Consider the eigenvectors $\{|x\rangle,|\bar{x}\rangle\}$, $\{|y\rangle,|\bar{y}\rangle\}$ and $\{|z\rangle,|\bar{z}\rangle\}$ of the Pauli operators $\sigma^x$ (X), $\sigma^y$ (Y) and $\sigma^z$ (Z), with eigenvalues $\{1,-1\}$, respectively. Then, the ground state of $H_{0}$ in (\ref{eqn:spinHam}) is given by
\begin{equation}
\begin{split}
|\phi_{0}\rangle=& \alpha|x_{1}x_{2}\bar{z}_{3}x_{4}x_{5}x_{6}\rangle+\beta|\bar{x}_{1}\bar{x}_{2}\bar{z}_{3}x_{4}x_{5}x_{6}\rangle \\
                 &+\mu|x_{1}x_{2}\bar{z}_{3}\bar{x}_{4}\bar{x}_{5}\bar{x}_{6}\rangle+\nu|\bar{x}_{1}\bar{x}_{2}\bar{z}_{3}\bar{x}_{4}\bar{x}_{5}\bar{x}_{6}\rangle,
\label{initial}
\end{split}
\end{equation}
where $\alpha$, $\beta$, $\mu$ and $\nu$ are complex amplitudes satisfying $|\alpha|^{2}+|\beta|^{2}+|\mu|^{2}+|\nu|^{2}=1$.
Experimentally, the ground state (\ref{initial}) of $H_0$ is represented as four spatial modes of single photons, as shown in the initial step of Fig.~\ref{setup}b. To evolve this state to the ground state of $H_1$ we only need to implement the additional ITE operations of $\sigma_{4}^{z}$ and $\sigma_{1}^{y}\sigma_{2}^{z}\sigma_{3}^{x}$. Particle 4 is expressed in the basis $\{|z\rangle,|\bar{z}\rangle\}$, as $|x_{4}\rangle = (|z_{4}\rangle+|\bar{z}_{4}\rangle)/\sqrt{2}$ and $|\bar{x}_{4}\rangle=(|z_{4}\rangle-|\bar{z}_{4}\rangle)/\sqrt{2}$. This change of basis transformation is implemented by half-wave plates (HWPs) in the initial four spatial modes. Eight spatial modes are created after splitting them by a BD30. The polarisation of the terms with $|z_{4}\rangle$, which represent states with higher energy, is set to be vertical with HWPs. The polarisation of the terms with $|\bar{z}_4\rangle$ is set to be horizontal.

The dissipative evolution is realised by passing photons through a polarisation beam splitter (PBS), where only four terms with horizontal polarisation remain at the end. Similarly, for the ITE of $\sigma_{1}^{y}\sigma_{2}^{z}\sigma_{3}^{x}$, the basis of particle 1 is rotated from $\{|x\rangle,|\bar{x}\rangle\}$ to $\{|y\rangle,|\bar{y}\rangle\}$ with the assistance of a combination of two HWPs and a QWP, as shown in Fig.~\ref{setup}f. Each of the spatial modes is horizontally split into two other modes with a BD30. For particle 2, the basis is rotated from $\{|x\rangle,|\bar{x}\rangle\}$ to $\{|z\rangle,|\bar{z}\rangle\}$. The eight spatial modes are further vertically split into sixteen modes with a BD60. The terms with the same form are combined with a BD30. Finally, the basis of particle 3 is changed to be $\{|x\rangle,|\bar{x}\rangle\}$, in which sixteen spatial modes are obtained with another BD30. After passing through a PBS, only the terms $\{|\bar{y}_{1}z_{2}x_{3}\bar{z}_{4}x_{5}x_{6}\rangle,$ $|\bar{y}_{1}\bar{z}_{2}\bar{x}_{3}\bar{z}_{4}x_{5}x_{6}\rangle,$ $|y_{1}\bar{z}_{2}x_{3}\bar{z}_{4}x_{5}x_{6}\rangle,$ $|y_{1}z_{2}\bar{x}_{3}\bar{z}_{4}x_{5}x_{6}\rangle,$ $|\bar{y}_{1}z_{2}x_{3}\bar{z}_{4}\bar{x}_{5}\bar{x}_{6}\rangle,$ $|\bar{y}_{1}\bar{z}_{2}\bar{x}_{3}\bar{z}_{4}\bar{x}_{5}\bar{x}_{6}\rangle,$ $|y_{1}\bar{z}_{2}x_{3}\bar{z}_{4}\bar{x}_{5}\bar{x}_{6}\rangle,$ $|y_{1}z_{2}\bar{x}_{3}\bar{z}_{4}\bar{x}_{5}\bar{x}_{6}\rangle\}$ remain and the output state corresponds to the ground state of $H_{2}$. The ITE of the other Hamiltonians that are part of the cyclic evolution are found in Section I.A of SM.

After the basis rotation shown in Fig.~\ref{setup}d the final state is expressed in the same basis as the initial state and takes the form
\begin{equation}
\begin{split}
|\phi_{4}\rangle &=(\alpha+\beta)|x_{1}x_{2}\bar{z}_{3}x_{4}x_{5}x_{6}\rangle+(\mu-\nu)|x_{1}x_{2}\bar{z}_{3}\bar{x}_{4}\bar{x}_{5}\bar{x}_{6}\rangle\\
&+(\beta-\alpha)|\bar{x}_{1}\bar{x}_{2}\bar{z}_{3}x_{4}x_{5}x_{6}\rangle+(\mu+\nu)|\bar{x}_{1}\bar{x}_{2}\bar{z}_{3}\bar{x}_{4}\bar{x}_{5}\bar{x}_{6}\rangle.\\
\end{split}
\label{final2}
\end{equation}

To clearly show the gate operation in the logical basis, we translate the basis by $|x_{1}x_{2}\rangle=(|0_{12}\rangle+|1_{12}\rangle)/\sqrt{2}$, $|\bar{x}_{1}\bar{x}_{2}\rangle=(|0_{12}\rangle-|1_{12}\rangle)/\sqrt{2}$, $|x_{4}x_{5}x_{6}\rangle=(|0_{456}\rangle+|1_{456}\rangle)/\sqrt{2}$ and $|\bar{x}_{4}\bar{x}_{5}\bar{x}_{6}\rangle=(|1_{456}\rangle-|0_{456}\rangle)/\sqrt{2}$. The logical basis is given by $|00_{g}\rangle = |0_{12}0_{456}\rangle|\bar{z}_{3}\rangle$, $|01_{g}\rangle = |0_{12}1_{456}\rangle|\bar{z}_{3}\rangle$, $|10_{g}\rangle = |1_{12}0_{456}\rangle|\bar{z}_{3}\rangle$ and $|11_{g}\rangle = |1_{12}1_{456}\rangle|\bar{z}_{3}\rangle$.
The initial state (ground state of $H_{0}$) is given in the logical basis by
\begin{equation}
\begin{split}
|\phi_{0}\rangle & = (\alpha+\beta-\mu-\nu)|00_{g}\rangle + (\alpha+\beta+\mu+\nu)|01_{g}\rangle \\
& \,\,\,\,+(\alpha-\beta-\mu+\nu)|10_{g}\rangle +(\alpha-\beta+\mu-\nu)|11_{g}\rangle.
\end{split}
\end{equation}
where, for simplicity, we omitted the overall normalisation.
After the anticlockwise braiding, the final state becomes
\begin{equation}
\begin{split}
|\phi_{4}\rangle & = (\beta-\mu)|00_{g}\rangle + (\beta+\mu)|01_{g}\rangle \\
&+(\alpha+\nu)|10_{g}\rangle +(\alpha-\nu)|11_{g}\rangle.
\end{split}
\end{equation}
The unitary transformation that corresponds to the anticlockwise braiding of MZMs A and C reads
\begin{equation}
U={1 \over \sqrt{2}}\left(
\begin{array}{cccc}
1&0&0&-1  \\
0&1&-1&0  \\
0&1&1&0  \\
1&0&0&1  \\
\end{array}
\right),
\end{equation}
written in the basis $\{|00_{g}\rangle, |01_{g}\rangle, |10_{g}\rangle, |11_{g}\rangle\}$.
If we focus on the even fermion parity sector spanned by $|00_g\rangle$ and $|11_g\rangle$, the unitary transformation becomes
\begin{equation}
U={1 \over \sqrt{2}}\left(
\begin{array}{cc}
1&-1  \\
1&1  \\
\end{array}
\right).
\end{equation}
As a result, the braiding of A and C corresponds to a generalised form of the Hadamard gate operation, related to the standard Hadamard gate by  $U\cdot\text{R}^2$.

{\em Experimental quantum process tomography.} In our experiment, we employ the quantum process tomography to identify the efficiency of the performed gate operations~\cite{OBrien2004}. The experimental measurement basis is chosen to be $\{hh, hv, vh, vv\}$, where $h$, $v$, $r$ and $d$ represent the horizontal, vertical, right-hand circular and diagonal polarisations, respectively. For each input state we need to reconstruct the final output state by two-qubit-state tomography with 16 measurement configurations, as shown in Fig. S13 of SM. To reconstruct the quantum process, we need 16 different input states. As a result, there are $16^2$ measurement settings. By expanding the output state $\mathcal{E}(\rho)$ in terms of the Pauli basis operators $\{\hat{E}_{m}\}=\{$${\rm II}$, ${\rm IX}$, ${\rm IY}$, ${\rm IZ}$, ${\rm XI}$, ${\rm XX}$, ${\rm XY}$, ${\rm XZ}$, ${\rm YI}$, ${\rm YX}$, ${\rm YY}$, ${\rm YZ}$, ${\rm ZI}$, ${\rm ZX}$, ${\rm ZY}$, ${\rm ZZ}$$\}$,
the quantum process can be expressed as $\mathcal{E}(\rho)=\sum_{mn}\chi_{mn}\hat{E}_{m}\rho{\hat{E}_{n}}^{\dag}$.
The physical process $\mathcal{E}$ is uniquely characterised by the 16-by-16 matrix $\chi$.

In our experiment, the spin basis is represented as $\{$$|x_{1}x_{2}\bar{z}_{3}x_{4}x_{5}x_{6}\rangle$, $|\bar{x}_{1}\bar{x}_{2}\bar{z}_{3}x_{4}x_{5}x_{6}\rangle$, $|x_{1}x_{2}\bar{z}_{3}\bar{x}_{4}\bar{x}_{5}\bar{x}_{6}\rangle$, $|\bar{x}_{1}\bar{x}_{2}\bar{z}_{3}\bar{x}_{4}\bar{x}_{5}\bar{x}_{6}\rangle$$\}$, which corresponds to the polarisation basis of $\{$$hh$, $hv$, $vh$, $vv$$\}$. The computation basis is chosen to be $\{$$|00_{g}\rangle$, $|01_{g}\rangle$, $|10_{g}\rangle$, $|11_{g}\rangle$$\}$. The transformation between the experimental basis and the computation basis is
\begin{equation}
\left(
\begin{array}{c}
|hh\rangle  \\
|hv\rangle  \\
|vh\rangle  \\
|vv\rangle  \\
\end{array}
\right)=U\left(
\begin{array}{c}
|00_{g}\rangle  \\
|01_{g}\rangle  \\
|10_{g}\rangle  \\
|11_{g}\rangle  \\
\end{array}
\right)={1 \over 2}\left(
\begin{array}{cccc}
1&1&1&1  \\
-1&1&-1&1  \\
1&1&-1&-1  \\
-1&1&1&-1  \\
\end{array}
\right)
\left(
\begin{array}{c}
|00_{g}\rangle  \\
|01_{g}\rangle  \\
|10_{g}\rangle  \\
|11_{g}\rangle  \\
\end{array}
\right).
\end{equation}
The output state in the computation basis can be represented as
\begin{equation}
\mathcal{E}(\rho')=\sum_{i,j}\lambda_{ij}\hat E_{i}\rho' \hat E_{j}^\dag.
\end{equation}
where
\begin{equation}
\lambda_{ij}=\sum_{n,m}\chi_{ij}\text{Tr}\big[(U^{\dag} \hat E_{n} U) \hat E_{i}\big]\text{Tr}\big[(U^{\dag} \hat E_{m} U)\hat E_{j}\big].
\end{equation}
A further restriction to the even parity sector can be performed by the projector $P_{e}=(|00_{g}\rangle\langle00_{g}|+|11_{g}\rangle\langle11_{g}|)/2$. This results in 4-by-4 reduced density matrices expressed in the logical basis $\{$$|00_{g}\rangle$, $|11_{g}\rangle$$\}$, as shown in Figs.~\ref{density} and \ref{noises}.

{\bf Acknowledgments}

{\bf Funding:} This work was supported by the National Key Research and Development Program of China (Grants NO. 2016YFA0302700), National Natural Science Foundation of China (Grant NO. 61725504, 11474267, 61327901, 11774335 and 61322506), Anhui Initiative in Quantum Information Technologies (Grant No.\ AHY060300 and AHY020100), Key Research Program of Frontier Science, CAS (Grant No.\ QYZDYSSW-SLH003), the Fundamental Research Funds for the Central Universities (Grant Number WK2470000020, WK2470000026 and WK2030380015), Youth Innovation Promotion Association and Excellent Young Scientist Program CAS and the UK EPSRC grant EP/I038683/1 and EP/R020612/1. {\bf Competing interests:} The authors declare that they have no competing interests. {\bf Author contributions:} Y.-J.H. proposed this project. J.-S.X. and K.S. designed the experiment. K.S. performed the experiment with the assistant of J.-S.X.. J.K.P. contributed to the theoretical analysis. J.-S.X., K.S., Y.-J.H. and J.K.P. wrote the manuscript. Y.-J.H. supervised the theoretical part of the project. C.-F.L. and G.-C.G. supervised the project. All authors read the paper and discussed the results. {\bf Data and materials availability:} All data needed to evaluate the conclusions in the paper are present in the paper and/or the Supplementary Materials. Additional data related to this paper may be requested from the authors.

\end{document}